%% file: main.tex
\documentclass[letterpaper]{article}

\usepackage{natbib,alifeconf}
\usepackage{graphicx}  
\usepackage{amsmath}
\usepackage{cleveref} 

\usepackage{breakcites}
\usepackage{array, caption, floatrow, tabularx, makecell, booktabs}%
\newfloatcommand{capbtabbox}{table}[][\FBwidth]
\setcellgapes{3pt}
\usepackage{subfiles}
\usepackage{times}
\usepackage{xcolor}
\usepackage{soul}
\usepackage[super]{nth}
\usepackage{geometry}
\usepackage[utf8]{inputenc}
\usepackage{textcomp}
\usepackage[utf8]{inputenc}
\usepackage{siunitx,booktabs}
\usepackage{times}  
\usepackage{helvet}  
\usepackage{courier}  
\usepackage{url}  
\usepackage{amssymb}

\graphicspath{ {eps/} }
\usepackage{booktabs}
\usepackage{latexsym} 
\usepackage{subcaption,siunitx,booktabs}
\usepackage{tikz}
\newcommand*\circled[1]{\tikz[baseline=(char.base)]{
            \node[shape=circle,draw,inner sep=1pt] (char) {#1};}}

\usepackage{caption}
\title{Pathways to Good Healthcare Services and Patient Satisfaction: An Evolutionary Game Theoretical Approach}
\author{Zainab Alalawi, The Anh Han, Yifeng Zeng \and  Aiman Elragig \\
\mbox{}\\
School of Computing and Digital Technologies, Teesside University, United Kingdom, TS1 3BX\\
Emails: \{z.alalawi, t.han, y.zeng, a.elragig\}@tees.ac.uk} 

\begin{document}
\maketitle

\begin{abstract}
 Spending by the UK\textquotesingle s National Health Service (NHS) on independent healthcare treatment has been increased in recent years and is predicted to sustain its upward trend with the forecast of population growth. Some have viewed this increase as an attempt not to expand the patients\textquotesingle~choices but to privatize public healthcare. This debate poses a social dilemma whether the NHS should stop cooperating with Private providers. This paper contributes to healthcare economic modelling by investigating the evolution of cooperation among three proposed populations: {\it Public Healthcare Providers}, {\it Private Healthcare Providers} and {\it Patients}. The Patient population is included as a main player in the decision-making process by expanding patient\textquotesingle s choices of treatment. We develop a generic basic model that measures the cost of healthcare provision based on given parameters, such as NHS and private healthcare providers\textquotesingle~cost of investments in both sectors, cost of treatments and gained benefits. A patient\textquotesingle s costly punishment is introduced as a mechanism to enhance cooperation among the three populations. Our findings show that cooperation can be improved with the introduction of punishment (patient\textquotesingle s punishment) against defecting providers. Although punishment increases cooperation, it is very costly considering the small improvement in cooperation in comparison to the basic model.
 
\end{abstract}


\input{introduction}
\input{relatedwork}
\input{method}

\input{result}
\input{conclusions}
\section{Acknowledgements}

This work was supported by EPSRC project No.\ (EP/S011609/1) and The Anh Han is supported by Future of Life Institute
(grant RFP2-154).

\bibliographystyle{apalike}

 \fontsize{7.8pt}{8.8pt}\selectfont
\bibliography{main} 

\end{document}

%% file: introduction.tex
\section{Introduction}
The NHS is a free healthcare service at the point of delivery in the UK funded through taxpayers\textquotesingle~contributions \citep{slawson2018NHS}. Many public healthcare services are currently allocating amounts of their budgets to sourcing services from Independent Healthcare Services Provider (ISP)/private sector.
Motivated by the NHS's {\it Five Year Forward View} (FYFV), we choose the patient as the core focus of healthcare planning who is to be included in the decision-making process \citep{inspections2014five,ham2017next} for better health, patient care and financial sustainability. Several systematic studies looking into the improvement of clinical decision-making find that most patients expect to be informed about their situation and the treatment required, and play an important role in their clinical decision-making. However, little attention has been given to understanding the patient\textquotesingle s role quantitatively as part of dynamic system modelling.  
 
To investigate the dynamic system interactions among Public and Private health sectors and Patients, we resort to an Evolutionary Game Theory (EGT)-based solution. Researchers have applied EGT in a wide range of disciplines running the gamut from economics, politics and security, to ecology, mathematical biology, and computer science \citep{adami2016evolutionary}. EGT allows us to understand and analyze the complex healthcare system, interactions between individuals from various populations in a game, and how strategic behavior might evolve among individuals \citep{nowak:ED,neumann:Togeb}. 

The main challenges of using EGT lie in formulating a valid payoff matrix and defining the ties between parameters within the proposed payoff matrix of each population. Some research on the evolution of cooperation has focused on the human willingness to engage in behavior that would involve paying a cost in return for imposing punishment on defectors or perceived wrongdoers \citep{sigmund2010social, dreber2008winners}.~Punishment is one mechanism that can enhance cooperation between individuals caught in social dilemmas \citep{hauert2007via}.~Punishment can be seen applied in human society, such as by punishing free-riders \citep{sigmund2010social}, and in governance or institutional systems that impose rewards and punishments on agents (participants) \citep{andreoni2003carrot}. Different types of punishment can be implemented based on the structure of the played game: peer punishments, pool punishments and institutional punishments \citep{sigmund2010social}.

In this article, we consider the patient\textquotesingle s role when developing an EGT decision-based model of healthcare services through a tripartite, one-shot EGT game \citep{gigerenzer2011h,elwyn2012shared}.~In our proposed model, we measure patients satisfaction and healthcare providers\textquotesingle~reputation which impact the quality of healthcare services \citep{ham2017next,robertson2017public}.~The model is aimed at identifying dynamic interactions that could enhance cooperation between the three populations.~It will allow us to ascertain how selecting certain decisions within a dynamic system would influence patients\textquotesingle~satisfaction with the provided service and their willingness to cooperate.~This could have important implications for addressing the significant, alarming drop in public satisfaction with NHS-provided services in recent years~\citep{robertson2017public}.~The social norm behavior of individuals is analyzed by applying EGT using one-shot game. 

The dynamic of the game is computed through stochastic selection of strategies based on patients\textquotesingle~satisfaction with received treatments and providers\textquotesingle~reputations.
The main question in this article is about how patients influence the dynamics of their cooperation in an EGT-based framework of three populations. The main contribution of this article is to build a basic model for the interactions of the three populations that closely capture the costs and benefits of every strategy combining decisions by agents in finite populations on either {\it cooperation} or {\it defection}. We further introduce peer punishment into the model: the patients\textquotesingle~punishment which takes the form of complaints for clinical negligence \citep{bryden2011duty, cooper:2010}.

The main contributions are summarized below. \begin{itemize}
\item We develop a simple, yet expressive, basic mathematical model to formulate interactions among the three populations: Patients, Private and Public health sectors. This is the first dynamical model that captures the interactions of these populations in healthcare economic modelling research; 
\item We develop a mechanism to study the behavior of individuals in each population and extend the basic model by introducing punishments in their interactions;
\item We analyze the models and examine how the relevant factors would influence cooperation among individuals within the populations;
\item We conduct a comprehensive simulation analysis to determine various types of behavior most-frequently adopted by individuals based on certain factors.
\end{itemize}


%% file: relatedwork.tex
\section{Related Work}\label{ssec:relatedwork}
The rapid development in research on the learning of social behavior has significantly increased our understanding of the dynamic interaction among individuals from different populations~\citep{nowak:ED,sigmund2010social}. Cooperation is one of the fundamental indicators to measure the strength and dynamism of a population~\citep{smith1974theory,kurokawa2009emergence,encarnaccao2016paradigm}. It can be studied by applying EGT using different types of mechanisms, such as reciprocal behaviors, mutual reciprocity among populations, replication, kin selection and costly punishment~\citep{nowak:five,hofbauer1998evolutionary}. Researchers seeking to understand the behavior of different agent representations within the healthcare system use AI~\citep{Anh2013WhyII}, game theory~\citep{Brekke2007}, multi-agent systems \citep{deSoham2017} and big data~\citep{murdoch2013inevitable} to predict and understand behaviors within the system.

\citep{Brekke2007} argued that having a blurred line between the private and public healthcare providers within the NHS might lead to imbalances in the costs of provided health services and a drift towards privatisation. While~\citep{wu2016game} developed their proposed set of various non-cooperative and cooperative games for the Emergency Department response based on different types of patients.
Another research investigates different dilemmas based on a three-population EGT framework involving the cost for prescribed antibiotics via healthcare providers~\citep{bettinger2016game}. The main limitations are related to the actual cost paid for prescriptions and efficiency in quantifying incentives of patients for selecting the most satisfying or preferred provider. Another research by  Encarna\c c\~ao \citep{encarnaccao2016paradigm} shows that the advent of the civil sector adds another layer of complexity to a scene that used to be dominated by two sectors: private and public.

Cooperation level is analysed by frequency-dependent selection of strategies within the populations~\citep{kandori1993learning, taylor2004evolutionary}. In this context, we seek to show how a drift towards a cooperate strategy in interactions between the three populations can be promoted
by the adoption of the most dominant strategy (social behavior) in our proposed model. Based on the stochastic factors and processes associated with the healthcare model, this article intends to investigate evolving societal behavior between patients and different sectors in the healthcare dynamic system. The selection of the patient population in our model was made for the following reasons: finding the best behavior and strategy for decision-makers among the three populations; determining the impact of implemented peer punishments by a patient and how this social mechanism could influence the decision-making process for better cooperation; and, finding the best strategy for involved populations by computing the interaction between private and public sectors~\citep{cooper:2010,Brekke2007}. 
A stochastic multi-objective auto-optimisation model was introduced by Bastian et al. to effectively manage  resource allocation for the military health system with an eye to achieving a more efficient funding and staffing distribution between the Army, Air Force and Navy~\citep{bastian2017stochastic}.
Bastian et al.\textquotesingle s research suffers from serious limitations; primarily, that the model introduced was not generic and depended on fixed inputs. 

The Patient population in our model plays a major role (as discussed in the results) in influencing decision-making. Our analysis significantly improves our understanding of the model structure and the factors that lead to cooperation. It helps us answer important questions such as: What factors might influence the patient\textquotesingle s rating of the healthcare services? What factors would induce healthcare providers to seek better reputations when the patient derives no or little benefit from the treatment (e.g. the patient files a complaint for receiving bad healthcare services).

%% file: method.tex
\section{EGT-Based Solutions}\label{sec:thealgorithm}

\subsection{Basic Model and Extended Model}\label{sec:Models}
\subsubsection{Model I - Basic model}\label{sec:Bthemodel}

In this model we consider three populations: Public providers, Private providers and consumers/Patients. While Public represents the NHS or the Public healthcare providers, Private independent healthcare providers sell healthcare services, and  Patient represents a person seeking treatment(s). An individual from each population (Public, Private and Patient) can choose from two strategies: provide/accept sustainable treatment(s) identified as cooperating, otherwise can't provide/refuse treatment(s) leading the patient to seek alternative treatment(s) from other providers. An agent\textquotesingle s payoff is acquired based on the strategy played by each individual from the three populations, as explained in Table~1.
{\small
\begin{table}[!t]
\begin{subtable}{\textwidth}
    \begin{tabular}{|p{5.6cm}|c|}
    \hline
    \textbf{Parameters\textquotesingle~description} &  \textbf{Symbol} \\
     \hline \hline
        Reputation benefit for the Public and Private healthcare providers & $b_R$ \\
      \hline
     Patient\textquotesingle s benefit & $b_P$\\
    \hline
     Cost of investment spent by the Public/Private healthcare provider & $c_I$\\
   \hline
    Cost of treatment acquired by the healthcare provider & $c_T$\\
   \hline
     Cost of healthcare management & $c_M$\\
     \hline
     Extra Patient\textquotesingle s benefit when both providers cooperate & $\varepsilon$\\
  \hline
 \end{tabular}
\end{subtable} 
\bigskip{\vspace{-0.3cm}}
\begin{subtable}{\textwidth}
\small
   \begin{tabular}{|p{0.2cm}|p{0.2cm}|p{0.2cm}||p{1.5cm}|p{1.6cm}|p{1.2cm}|}
   \hline
   \multicolumn{3}{| p{0.6cm} ||}{\textbf{Strategies}} &  \multicolumn{3}{ c |}{\textbf{Payoffs}} \\
    \hline
   \textbf{P1} & \textbf{P2} & \textbf{P3} & \textbf{Public} & \textbf{Private} & \textbf{Patient} \\
   \hline \hline
    \textit{C} & \textit{C} & \textit{C} & $b_R-c_I-c_T$ & $ b_R-c_I-c_M$ & $b_P+\varepsilon b_P$ \\
   \hline
    \textit{C} & \textit{C} & \textit{D} & $-c_I$ & $-c_I$ & 0\\
   \hline
    \textit{C} & \textit{D} & \textit{C} & $b_R-c_I-c_T$ & 0 & $b_P$\\
   \hline
    \textit{C} & \textit{D} & \textit{D} & $-c_I$ & 0 & 0 \\
   \hline
    \textit{D} & \textit{C} &\textit{C} & 0 & $b_{R}-c_I-c_M$ & $b_P-c_T$\\
  \hline
    \textit{D} & \textit{C} & \textit{D} & 0 & $-c_I$ & 0 \\
  \hline
   \textit{D} & \textit{D} &\textit{C} & 0 & 0 & $-c_T$\\
  \hline
    \textit{D} & \textit{D} & \textit{D} & 0 & 0 & 0\\
 \hline
\end{tabular}\end{subtable}
\normalsize
\caption{The healthcare model (Public healthcare providers \textit{P1}, Private healthcare providers \textit{P2} and Patient \textit{P3}).}
\label{tbl:BModel}
\end{table}
}

Every individual or agent in each of the three populations experiences one of the following scenarios based on two strategies, namely, \textit{cooperate (C)} and \textit{defect (D)}. 
This allows us to understand how cooperation evolves in altruistic interactions among individuals in a game. 
The following are the strategies an individual within each population can select (refer to Table~1):\\  
\textbf{Public healthcare providers}:~\circled{1} (Cooperate, \textit{C})~offers treatment paid for from taxpayers\textquotesingle~money, 
in return gets a reputation benefit. (In case of cooperation, the public commits to invest~($c_I$)~from allocated budget). \circled{2} (Defect, \textit{D})~does not want to pay for the treatment.\\
\textbf{Private healthcare providers}:~\circled{1} (Cooperate, \textit{C})~offers treatment either paid by Public
 (when Public cooperates) or self-paid by Patient (so the main cost involved is represented by management cost ($c_M$)), 
 and obtains a reputation benefit~($b_R$). In case of cooperation with the Patient, Private commits to invest ($c_I$) 
 from its revenue. \circled{2} (Defect, \textit{D})~does not want to offer the treatment.\\
\textbf{Patient}:~\circled{1} (Cooperate, \textit{C})~accepts the treatment and pays for the treatment~($c_T$)~in the Private instance; Patient obtains health benefit~($b_P$) if one provider \textit{cooperates} and an extra health benefit ($\varepsilon b_P$) when both providers \textit{cooperate}. \circled{2} (Defect, \textit{D})~rejects the treatment and looks for alternative treatment mostly overseas.\\
 
The main issue we investigate in this article is the spending and cost effectiveness in the healthcare system with 
an eye to elucidating the social dilemma as mapped in Table~1. The dilemma questions the probability of cooperation (C) among the two sectors (Public and Private) and consumers (Patient) resulting in sustainable spending and cost effectiveness of the provision of treatment funded from the taxpayers\textquotesingle~money\footnote{O\textquotesingle Connell in this https://bit.ly/2VBY4im, article talks about finding a structural change to reduce NHS spending and enhance the tendency towards saving.}.

When applying the evolutionary rules as explained in Fig.~\ref{fig:BModelRules} on the matrix given in Table~1, 
the payoff is a simple representation of the healthcare cost-effectiveness, healthcare providers\textquotesingle~reputation benefit and patient\textquotesingle s benefit. The eight possible strategic scenarios are described below:

\begin{figure}[!t]
\includegraphics[width=\textwidth]{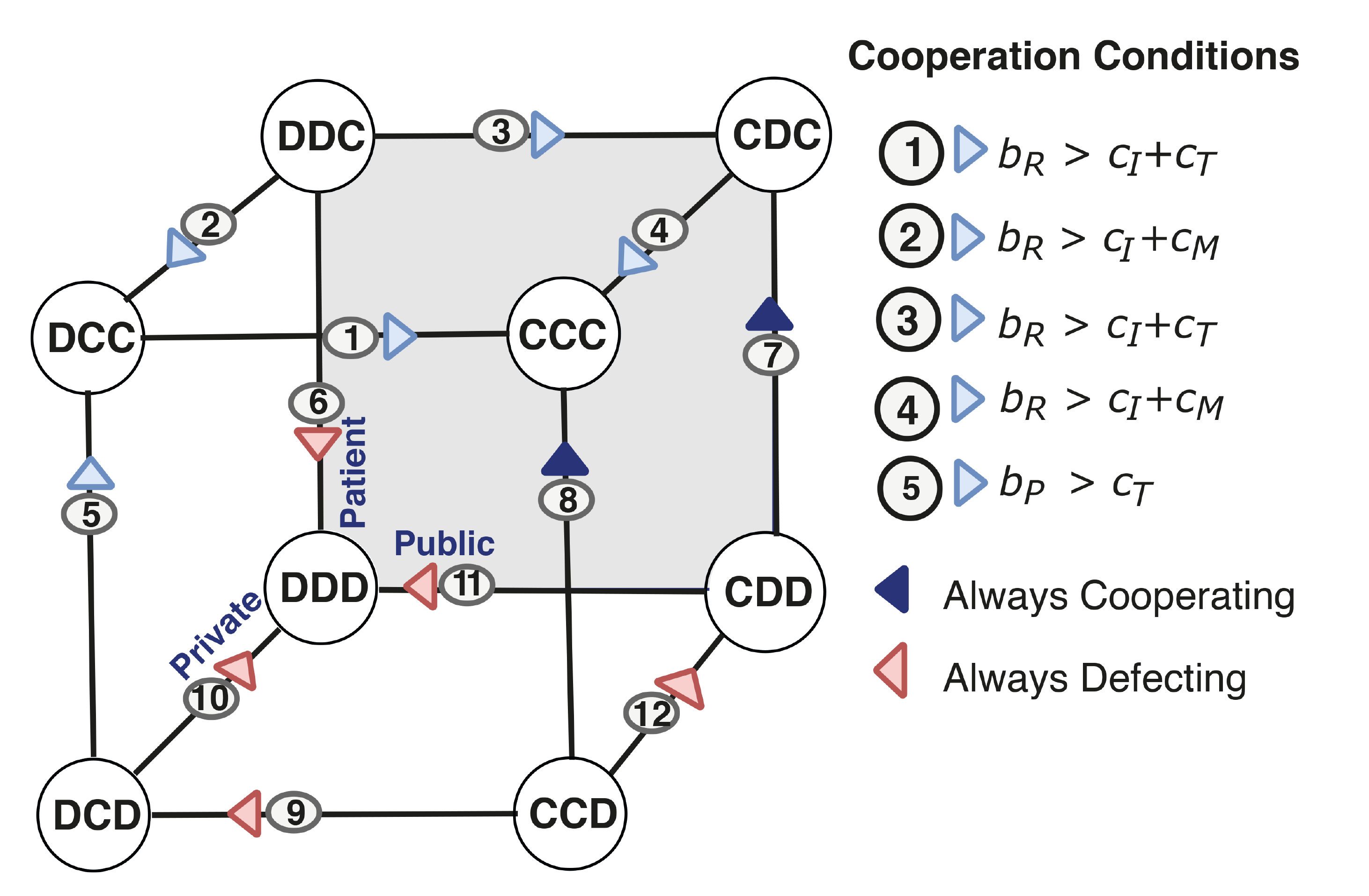}
\caption{Evolution dynamics of the simplex\textquotesingle s edges represents three-player actions (Basic model).}
\label{fig:BModelRules}
\end{figure}

\textbullet~Individuals from all three populations choose to \textit{cooperate, (CCC)}. In this case, the Public 
opts to pay for the treatment provided by the Private sector and the Patient accepts the provided treatment in pursuit of her/his own benefit of wellbeing and better health. The Public bears the costs of investment and pays for the Patient\textquotesingle s treatment ($c_I+c_T$) from its allocated budget. It gains reputation benefits ($b_R$) derived from the treatment provided by the Private healthcare provider. Reputation benefits are derived from the Patient\textquotesingle s satisfaction with the provided service. 
On the other hand, the Private sector will provide the required treatment to the Patient and receive the payment covering the costs from the Public. The cost of investment ($c_I$) of the Private healthcare provider is to invest in staffing and healthcare facilities, while ($c_M$) includes administrative and operational costs. Consequently, both the Private  and the Public  obtain a reputation benefit ($b_R$). Patient obtains extra health benefit ($\varepsilon b_P$) as both healthcare providers are cooperating, where ($\varepsilon$) captures a fraction of the Patient benefit.

\textbullet~Both the Private and Public healthcare providers want to pay and provide treatment to the patient, but the patient rejects the service~(\textit{CCD}). The payoff indicates that the Public healthcare provider will still invest ($c_I$) back in the Public as it is set to cooperate. Similarly, the Private sector only invests ($c_I$) back into its own resources, which include staffing, equipment and research, and gets nothing in return. The Patient\textquotesingle s payoff is 0 as no treatment cost was involved; nor did she/he get health benefits from the services of the healthcare providers involved.
{\small
\begin{figure}[!t]
\includegraphics[width=\textwidth]{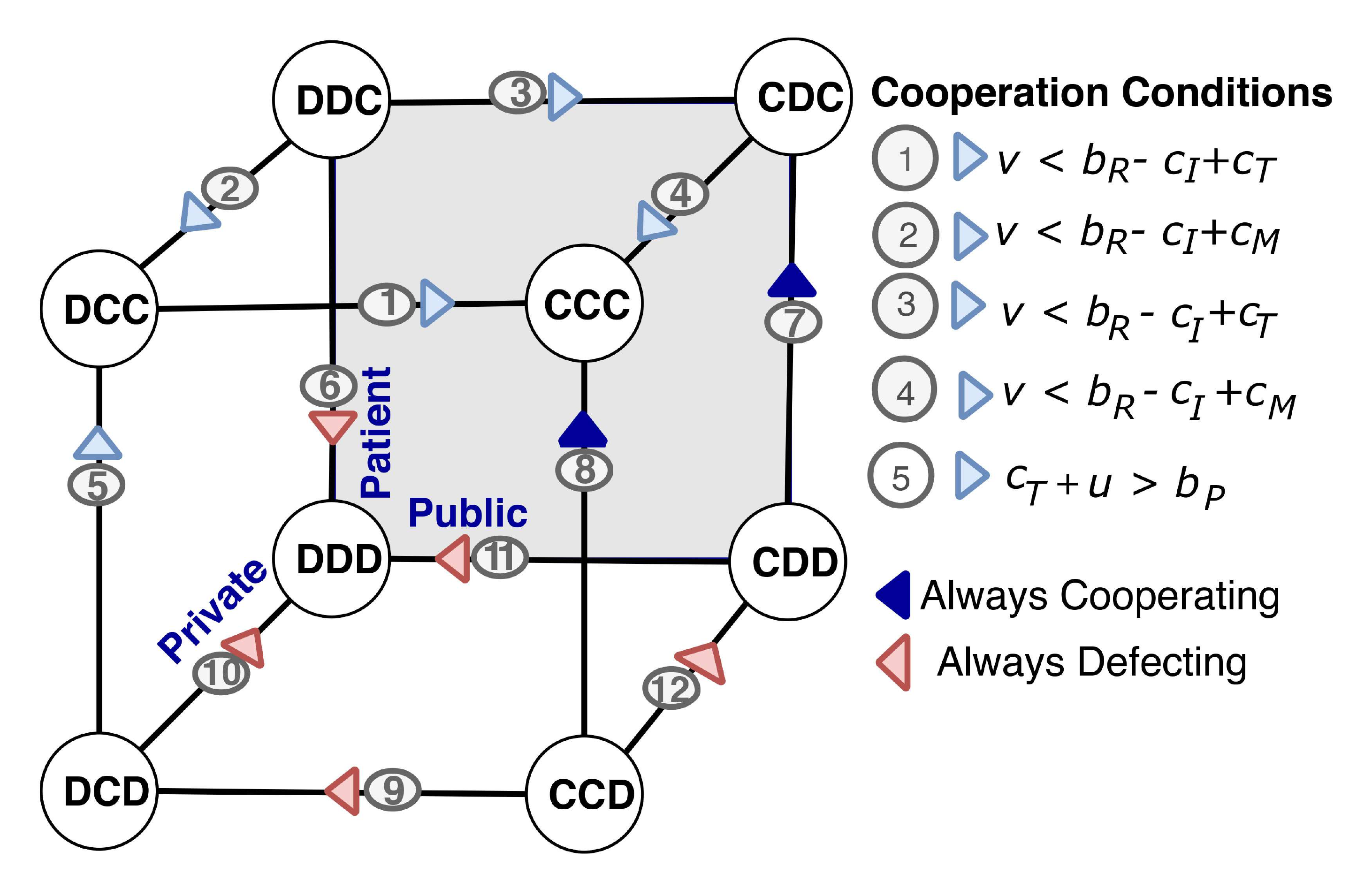}
\caption{Evolution dynamics of the simplex\textquotesingle s edges represents three-player actions (Extended model with patient\textquotesingle s punishment). }
 \label{fig:PModelRules}
\end{figure}}

\textbullet~The Patient accepts the treatment to be provided by the Private sector and paid for by the Public but the Private refuses to provide the treatment~(\textit{CDC}). The Public\textquotesingle s payoff consists of the cost of investment and treatment ($c_I+c_T$), which are covered from its allocated budget, and a reputation benefit ($b_R$) is accumulated based on patients\textquotesingle~satisfaction. The cost of accepting the treatment for the Patient is 0 and s/he obtains health benefit ($b_P$).

\textbullet~The Public is the only party willing to provide treatment, but the Patient rejects the treatment offered~(\textit{CDD}). So, there is neither treatment cost nor benefit returned. Hence, the payoff for both the Private healthcare provider and the Patient is 0, while the Public still bears a cost of investment ($c_I$) in new treatments \citep{sapina2018protons}.

\textbullet~The Public healthcare provider opts out of providing treatment and refuses to pay for it; therefore, the payoff for the Public is 0 (\textit{DCC}). As the Public does not want to provide and pay for the Patient\textquotesingle s treatment, the Patient would look for treatment provided by a Private sector in a competitive market. The Private sector\textquotesingle s payoff is derived from the gains it makes in its own reputation benefit ($b_R$); even when it is not cooperating with the Public. The payment received goes towards defraying the costs of investment and management ($c_I+c_M$). The Patient, on the other hand, gets health benefits and pays the cost of the treatment ($b_P - c_T$).

\textbullet~Only the Private healthcare provider offers to provide treatment with a specified price tag while neither the Public wants~
to pay for the treatment nor the Patient is accepting the treatment~(\textit{DCD}). The payoff for the latter two~
agents is 0, while the transaction involves an investment cost ($c_I$) to be taken out of the Private provider\textquotesingle s budget.

\textbullet~None of the healthcare providers is willing to provide treatments~(\textit{DDC}). This situation leads the Patient~
to look for alternative treatments, both domestically and possibly overseas, and certainly the Patient has to pay for a treatment cost ($c_T$).

\textbullet~The situation depicts all three agents choosing not to interact with one another~(\textit{DDD}). Hence, there will be no winner or loser and all agents get a 0 payoff.

\begin{table}[!t]
\begin{subtable}{\textwidth}
    \begin{tabular}{|p{5.8cm}|c|}
    \hline
    \textbf{Parameters\textquotesingle~description} &  \textbf{Symbol} \\
     \hline \hline
 	 Patient\textquotesingle s punishment cost & v\\
 	 \hline
 	 Fee paid by punished agent & u\\
  	\hline
     \end{tabular}
\end{subtable}
\bigskip{\vspace{-0.3cm}}
\begin{subtable}{\textwidth}
\small
      \begin{tabular}{|p{0.2cm}|p{0.2cm}|p{0.2cm}||p{1.5cm}|p{1.6cm}|p{1.4cm}|}
   \hline
   \multicolumn{3}{| p{0.6cm} ||}{\textbf{Strategies}} &  \multicolumn{3}{ c |}{\textbf{Payoffs}} \\
    \hline
    \textbf{P1} & \textbf{P2} & \textbf{P3} & \textbf{Public} & \textbf{Private} & \textbf{Patient} \\
    \hline \hline
     \textit{C} & \textit{C} & \textit{C} & $b_R-c_I-c_T$ & $b_R-c_I-c_M$ & $b_P+\varepsilon b_P$ \\
    \hline
     \textit{C} & \textit{C} & \textit{D} & $-c_I$ & $-c_I$ & 0\\
    \hline
     \textit{C} & \textit{D} & \textit{C} & $b_R-c_I-c_T$ & -v & $b_P-u$\\
    \hline
     \textit{C} & \textit{D} & \textit{D} & $-c_I$ & 0 & 0 \\
    \hline
     \textit{D} & \textit{C} &\textit{C} & -v & $b_{R}-c_I-c_M$ & $b_P-c_T-u$\\
   \hline
     \textit{D} & \textit{C} & \textit{D} & 0 & $-c_I$ & 0 \\
   \hline
    \textit{D} & \textit{D} &\textit{C} & -v &-v& $-c_T-2u$\\
   \hline
     \textit{D} & \textit{D} & \textit{D} & 0 & 0 & 0\\
  \hline
 \end{tabular}
 \end{subtable}
\caption{The healthcare model with Patient\textquotesingle s Punishment (Public healthcare providers \textit{P1}, Private healthcare providers 
\textit{P2} and Patient \textit{P3}).}
\label{tbl:PModel}
\end{table}

\subsubsection{Model II- Model with Patient\textquotesingle s Punishment}\label{sec:Pthemodel}

This model extends the basic version by introducing peer punishment. Patient has the choice to punish defected healthcare providers after a one-shot game has been played. Costly punishment means the patient pays a cost \textit{u} to force the defecting healthcare provider to pay a cost of clinical negligence \textit{v}. The model payoff matrix is provided in Table 2. (Here we assume that, \textit{u} $<$ \textit{v})~\citep{fowler2005altruistic}.

When an individual from healthcare providers fails to fulfill their commitment to provide acquired healthcare services up to standards, a patient\textquotesingle s punishment would be meted out to the defecting individual(s) from healthcare providers \textit{v}; simultaneously a cost \textit{u} is accrued from the patient in this process. Furthermore, if a co-player refuses to commit, the payoff for both is 0. The transition probability of a mutant playing strategy D to invade a population of C players can be measured following the method explained below.

\subsection{Method: Evolutionary Dynamics for Three Populations}\label{ssec:evolutionaryDynamics}
EGT method is adopted  to study the evolutionary dynamics and interactions among individuals from three distinct finite populations: Public healthcare providers \textit{P1}, Private healthcare providers \textit{P2} and Patient \textit{P3}. Here, we assumed that the populations are of a fixed size \textit{N}. Every individual in each population will be involved in one of the eight strategic scenarios as mentioned earlier. Individuals have the choice to \textit{cooperate,~(C)} or to \textit{defect,~(D)} in a paradigm shift fashion. 
In our proposed model, there are eight possible paradigms corresponding to the the eight possible combinations of the basic strategies within the three populations, namely, \textit{CCC, CCD, CDC, CDD, DCC, DCD, DDC} and {\it DDD}. Denoting the number of cooperators in P1, P2 and P3 by $x$, $y$, and $z$, respectively, the payoff of each strategy can be  written  as follows:~
\begin{eqnarray}\label{eq:3.2}
\left.\begin{array}{rcl}
P_s^{Public}(x,y,z) &=& P_{syz}\\
P_s^{Private}(x,y,z) &=& P_{xsz}\\
P_s^{Patient}(x,y,z) &=& P_{xys}, \\
\end{array}
\right\}\qquad 
\end{eqnarray}
{\scriptsize
\begin{figure}[!t]
\includegraphics[width=\textwidth]{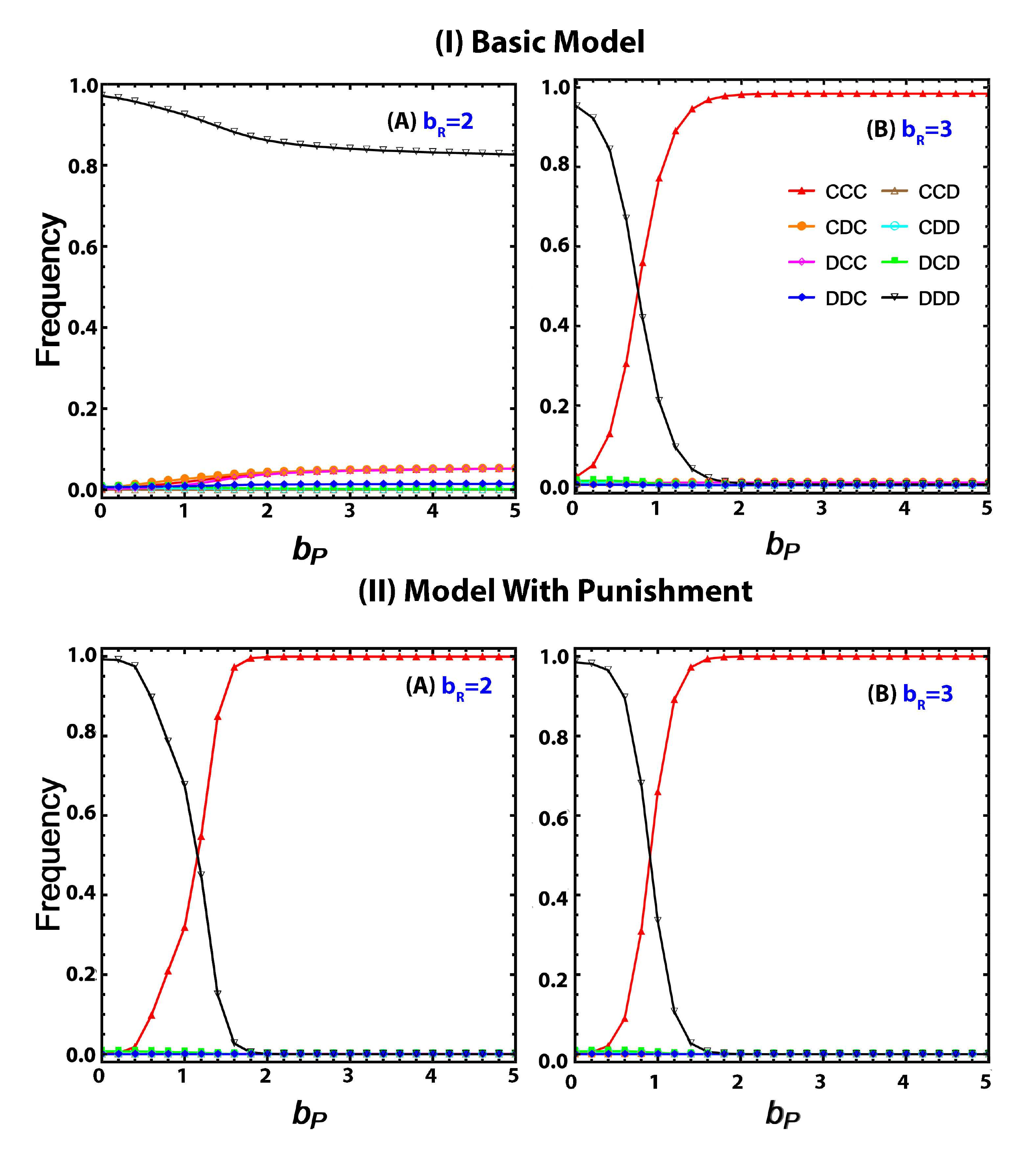}
\caption{The plots represent the frequency of all strategies adopted by the populations. Panel~(I) represents the frequency of all strategies adopted by the populations based on the basic model\textquotesingle s matrix. Panel~(II) examines the frequency of the model\textquotesingle s with punishment strategies for varying ($b_P$ and $b_R$) as stated, where: $u=0.5$ and $v=1.5$. Other parameters:~$c_I,~c_T,~c_M=1,~b_R=2$ and  $3,~\varepsilon=0.2,~N=100$~and~$\beta=0.1$.}
\label{fig:Frequencies}
\end{figure} 
}
where $x,y,z\in \{0,1\}$, is the payoff for the strategy selected by individuals from one of the stated populations, and (\textit{x,y,z}) represents the selected strategies \textit{C} or \textit{D}. For instance, individuals from \textit{P1} have the options to play \textit{C} or \textit{D}. The selected strategy will replace the \textit{s} at \textit{x} vertex, while \textit{y} and \textit{z} vertices remain unchanged for every selected strategy for the Public population in this instance.
The payoff of randomly selected individuals \textit{A} and \textit{B} in the population depends on the proportion of both players in the population.
In each time step an individual \textit{B} with fitness $\pi_B$ imitates a randomly selected individual \textit{A} with $\pi_A$ fitness adopting a pairwise comparison rule. The  probability $\rho$ that A adopts B\textquotesingle s strategy is given by the Fermi\textquotesingle s function \citep{sigmund2010social, traulsen2007pairwise}
\begin{eqnarray}\label{eq:3.1}
\rho = {[1+ e^{-\beta[\pi_A - \pi_B]}]}^{-1}
\end{eqnarray}
where the parameter $\beta\geq 0$ represents the \lq intensity of selection\rq or \lq imitation strength\rq ($\beta = 0$ represents neutral drift where imitation decision is random, while for large $\beta \rightarrow \infty$ the imitation decision is increasingly deterministic).
\begin{figure}[!t]
\includegraphics[width=\textwidth]{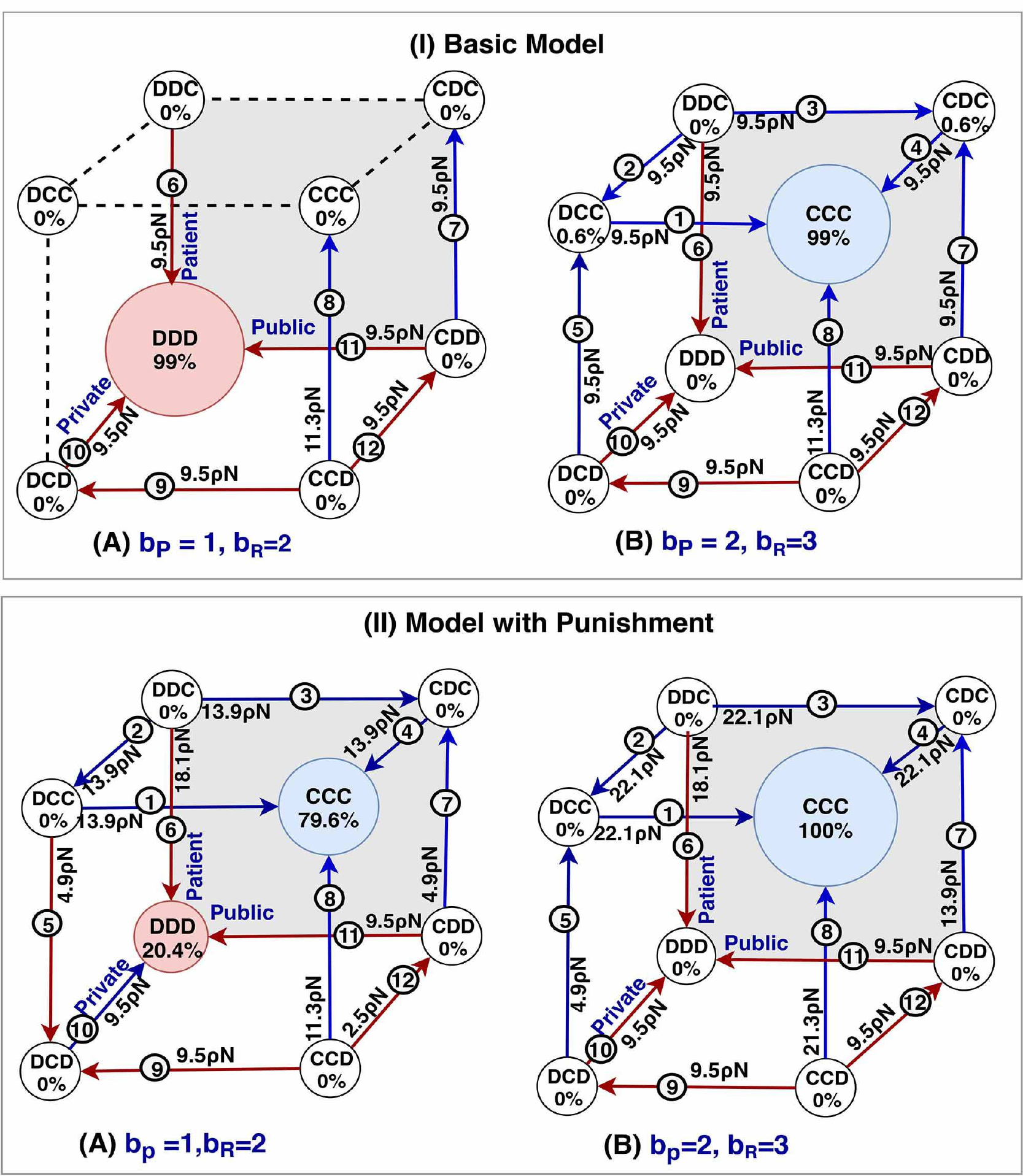}
\caption{Stationary distribution and fixation probabilities. The parameters' values are stated for each simplex, whereas: red arrow represents transition towards defection, blue arrow transition towards cooperation and dashed line refers to neutral state. A high fraction 80\% of individuals would adopt cooperation with lower $b_P$ in (II-A) compared to (I-A), whilst an adequate 1\% increase in cooperation with higher $b_R$ in (II-B) compared to (I-B) is registered. The transition probability and frequency dependency are normalised ($1/N$), where $N=100$. Other parameters: $c_I, c_T, c_M=1, \varepsilon=0.2$ and $\beta=0.1$.}
\label{fig:SDResult}
\end{figure}
In order to construct a symmetric matrix, the fitness of an individual adopting a strategy \textit{s} within a population is derived from the average obtained from the tripartite one-shot game described in Table~1. 
The social dynamics of the three finite populations interacting in eight strategies as a combination of \textit{C}s and \textit{D}s is represented by a death-birth process using pairwise comparison \citep{traulsen2006stochastic,nowak:ED}. 
Individuals with the highest payoff reproduce and their social behavior is adopted by weak opponents, i.e. the invading agent. For instance an agent \textit{B} from the Public population imitates the successful strategic behaviour, \textit{C}, by another regional healthcare provider as agent \textit{A}. Subsequently, the transition matrix is evaluated by values given to the associated parameters as depicted in Table~1.
At each time step, there is a probability of stochastic selection of an individual from a population whereby an individual \textit{B} from one of the populations playing \textit{D} with payoff $f_D$ may imitate another randomly selected individual \textit{A} with
payoff $f_C$ from the same population. The probability of the occurrence of this action \citep{nowak:ED,kandori1993learning} is stated in the equation above. The transition probability drifts for agent \textit{A} playing \textit{C} from \textit{k} to $k^{\pm}$ is given by:
\begin{eqnarray}\label{eq:3.5}
T^{\pm}(k) =\frac{k}{N}\frac{(N-k)}{N}[1+ e^{\mp \beta(\Pi_C(k)-\Pi_D(k))}]^{-1}
\end{eqnarray}

As mentioned earlier, the process has two absorbing states $k=0$ and $k=N$. In mixed populations, one of the absorbing states at the end would be a population with either all-\textit{C} or all-\textit{D}. Determining the different fixation probabilities $\rho_{D,C}$ is given by: 
\begin{eqnarray}\label{eq:3.6}
\rho_{D,C}=\left(1 + \sum_{i=1}^{N-1}\prod_{j=1}^i\frac{T^-(j)}{T^+(j)}\right)^{-1}
\end{eqnarray}
The transition matrix $\Lambda$ with a set of $\left\{1,\ldots ,s\right\}$ strategy \citep{encarnaccao2016paradigm,nowak:ED,Anh2013WhyII} is:
\begin{eqnarray}\label{eq:3.7}
\Lambda_{ij,j\neq i}=\frac{\rho_{ji}}{3}\hspace{1mm} \text{ and } \hspace{1mm} \Lambda_{ii}= 1- \sum_{j=1,j\neq i}^s \Lambda_{ij}
\end{eqnarray}
The various fixation probability acquired from $\rho_{ij}$ is that a population at a single state $i$ transits to another state $j$ when a mutant from one of the populations adopts an \textit{s} different strategy.
In games with large N, an invader can emerge as the stronger if the condition below is correct \citep{sigmund2010social}:
\begin{eqnarray}\label{eq:3.8}
\sum_{i=1}^{N-1}\Pi_{C}(k)  > \sum_{i=0}^{N}\Pi_{D}(k)
\end{eqnarray}

%% file: result.tex
\section{Results}\label{sec:resul}

\subsection{Pathways to healthcare cooperation}\label{ssec:result1}

\begin{figure}[!t]
\includegraphics[width=\textwidth]{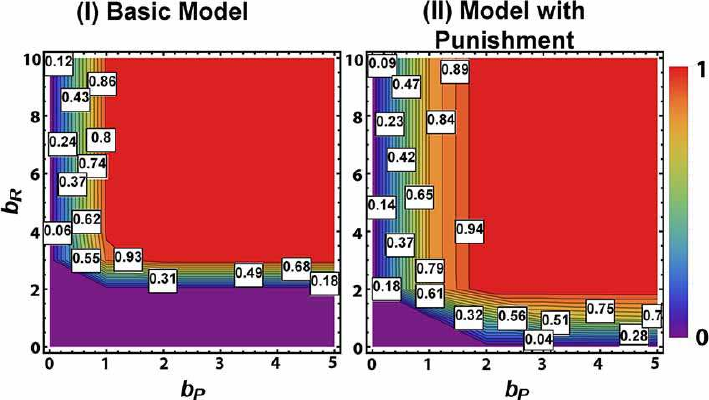}
\caption{Frequency of strategy \textit{CCC} for varying the main parameters $b_R$ and $b_P$:~(I) for the basic model, (II) extended model with costly punishment. In (II) $u =0.5$ and $v=1.5$. In general, \textit{CCC} performs better when punishment is introduced to the basic model. Additionally, significant cooperation is achieved for sufficiently low $b_R$ and large $b_P$. Other parameters $c_I,c_T,c_M =1, \varepsilon=0.2, N=100,$ and $\beta=0.1$.}
\label{fig:ControuPlot1}
\end{figure}

In conducting a numerical and systematic analysis for the basic model, we focus on the social interactions between players in each population and how their decision influences the level of cooperation to sustain cost-effective services and better patient satisfaction.~(Fig.~\ref{fig:Frequencies}~I) represents the computation investigating the frequency of adopting one of the eight strategies in the basic model by analysing the stochastic behavior of mutation in one of the three populations based on the frequency for each of the eight strategies given in (Table~1). That allows us to measure the ultimate behavior of those adopting the same strategy following the rules of social learning \citep{rendell2010copy}.

In the basic model, refer to (Fig.~\ref{fig:Frequencies}~I-A), where $b_R$ is small ($b_R$ = 2), the defection strategy \textit{DDD} is pervasive. The players of each of the three populations spend most of their time adopting defecting strategies rather than cooperating strategy. By simulating the matrix implementing (Eqs.~\ref{eq:3.6} and \ref{eq:3.7})  with a selected range of examined parameters ($b_P$ and $b_R$), as stated in (Fig. \ref{fig:SDResult} I-A), the analysis shows that the \textit{DDD} strategy dominates the populations' dynamic by 99\%.  
In other words, healthcare providers have to invest more efforts in order to satisfy patients as better cooperation \textit{CCC} is achieved when ($b_R$) is sufficiently high.

\begin{figure*}[!t]
\includegraphics[width=\textwidth]{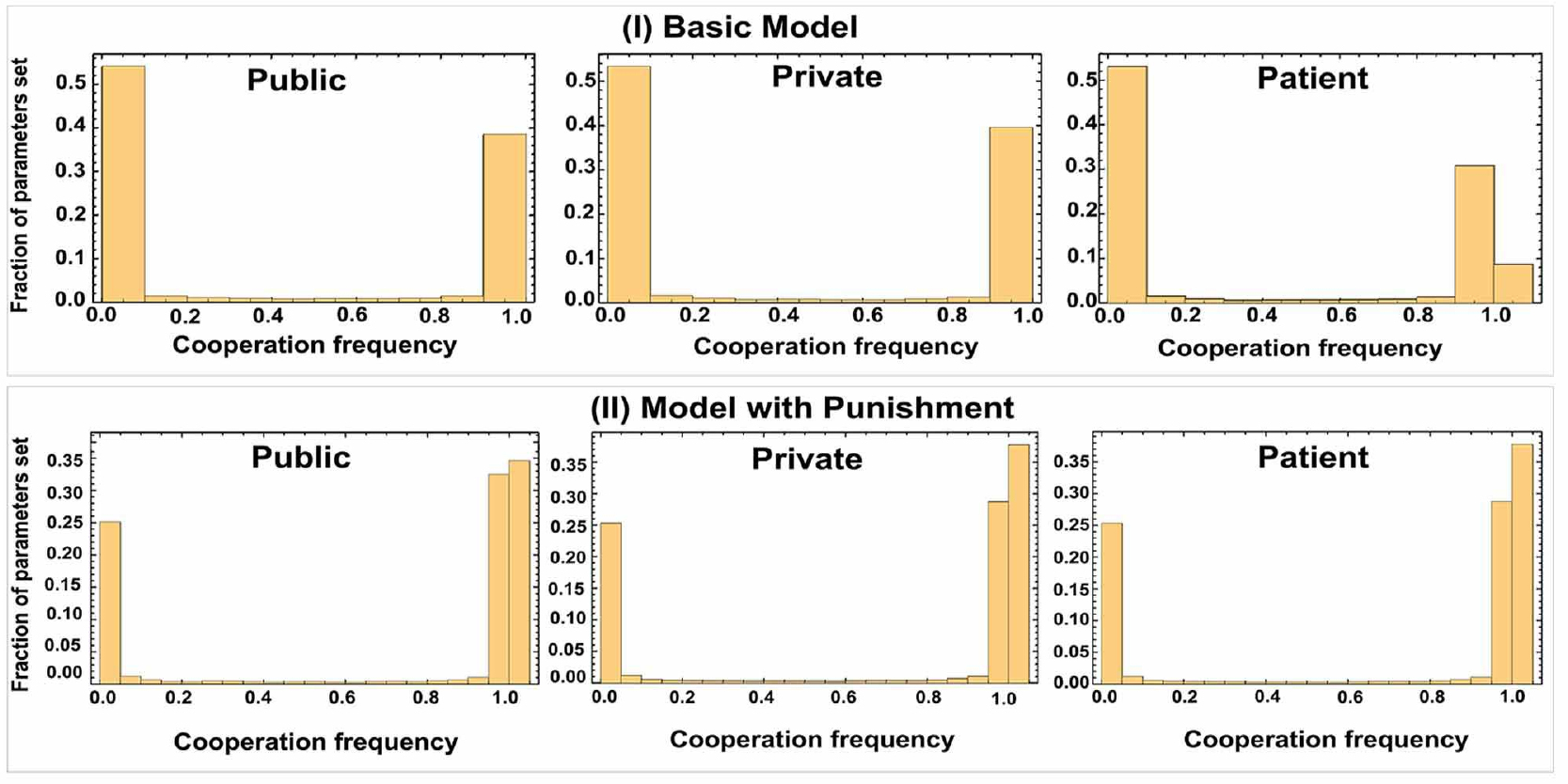}
\caption{Robustness of the results across game configurations and parameters of the model with or without punishment (using 10000 samples). For the basic model: (I) Parameter values: $0.5\leq b_P \leq 5$, $0.5\leq b_R\leq5$. (II) the model with punishment, where  parameter values: $0.5\leq b_P\leq5$, $0.5\leq b_R\leq5$, $0\leq u\leq3$ and $u\leq v\leq10u$ randomly sampling from uniform distributions on the intervals. Other parameters: $c_M,c_I,c_T=1$ and $\varepsilon=0.2, N=100,\\\beta=0.1$.}
\label{fig:Histogram}
\end{figure*}

As it has been observed in the basic model where punishment is absent, players of each population spend most of their time at the defecting strategy (see Fig.~\ref{fig:SDResult} I-A). We started by pairwise computation of the interaction strategies in the payoff matrix (Table~2) based on different values of the parameters ($b_R$ and $b_P$) to measure the stationary distribution and the frequency of the eight strategies. Recalling that in our model the patient has the option to mete out a costly punishment \textit{v} to the defecting healthcare provider(s) at~\textit{u} cost (i.e. legal fees).

Most finite populations drift towards cooperation (i.e, \textit{CCC}) as the most dominant strategy. In the presence of patient\textquotesingle s punishment (see Fig.~\ref{fig:SDResult}~II-A~$\&$~B), cooperators invade defectors when there is a large enough patient\textquotesingle s benefit (namely, $b_P > 1$), where the populations spend at least 79.6\% of their time in cooperation. Our simulation suggests that, regardless of the reputation\textquotesingle s benefit, cooperation is highly achievable (see Fig. \ref{fig:SDResult} II).

\subsection{Robustness of Parameters}\label{ssec:robustnessOfParameters}

A more compelling analysis has been carried out by computing 10000 samples with the set of parameters stated in (Fig.~\ref{fig:Histogram}) to obtain the stationary distribution of cooperation for both the basic model and the extended model. Considering the collected results in the plot for the basic model in (Fig.~\ref{fig:SDResult}), it is noticeable that the mean outcomes in (Fig.~\ref{fig:Histogram} I$\&$II) are approximate. 


%% file: conclusions.tex
\section{Discussion and Conclusions}\label{ssec:Discussion}
{\small

In summary, this paper investigates the behavior of individuals within the three finite populations to achieve better cooperation.~By analyzing the performance of each population and its preferred path towards evolving and adopting a new strategy, we explored how to move towards cooperation while taking into consideration the cost of effectiveness and patient satisfaction with the provided services.~Basically, the cooperation rules of each of the model\textquotesingle s dynamics explained in (Fig.~\ref{fig:BModelRules} \& \ref{fig:PModelRules}) clearly show that there are some paths moving toward full cooperation (see the panels in Fig.~\ref{fig:BModelRules}).

In contrast, avoiding the enveloped state (where some critics alluded to great fear of collaboration with private healthcare providers as \lq the beast of the (P-Privatisation)\rq) is represented by the \textit{DDD} node; 
the public sector represents the main agent that can take the lead in changing the rules and influence the behavior of the private sector and patient by introducing new policies.
For different values of $b_R$ and $b_P$, the frequency of strategy \textit{CCC} has improved in the extended model (as apposed to the basic model), (see Fig.~\ref{fig:ControuPlot1} A $\&$ B).

Moreover, the initial result yielded by the basic model is generally moving towards defection, apart from cases where a considerably high reputation value is imposed on the computation process, in which case there is a drift towards defection.
In comparison, the results obtained when patient's costly punishment is introduced for the populations show a heightened tendency towards cooperation. This requires not only high patient\textquotesingle s satisfaction but also a high reputation gained by healthcare providers; however, a full cooperation can be achieved within the basic framework on considerably low patient's satisfaction. In this case, the imposed peer punishment enhances cooperation with low patient\textquotesingle s satisfaction, but it is proved too costly to be adopted in the proposed healthcare model. 
Realistic scenarios could be derived from some elective treatments such as hip replacements \citep{moscelli2016market}. In fact, data collected by the NHS and Care Quality Commission (CQC) \citep{cqcRating}, and The Patient Reports Outcome Measures (PROMs) \citep{PROM} are well represented in our proposed model in addition to some NHS statistics that substitutes for the cost of treatment and expenditures.
We do recognise that our proposed models have some limitations, such as that our models are not validated against real healthcare data (it will be considered in future work) and the lack of comparison with other techniques. In this work, we only focus on analysing the behavior between agents using basic factors.\\
Future work will involve investigating the implementation of institutional punishment, expanding the model (i.e., by introducing new factors) to study the relation between the new elements and the dynamic behaviour of individuals.~Finally, our proposed model directs attention to how the patient's decision would impact the process of collaboration between healthcare providers, and to the effectiveness of management decisions made by the private sector in influencing the patient\textquotesingle s choice of cooperation.

}